\documentclass[12pt]{article}

\usepackage{scicite}
\usepackage{times}
\usepackage{graphicx}
\usepackage{enumitem}
\newenvironment{sciabstract}{%
\begin{quote} \bf}
{\end{quote}}

\newcommand{\beqar}{\begin{eqnarray}}
\newcommand{\eeqar}{\end{eqnarray}}

\newcommand{\bcen}{\begin{center}}
\newcommand{\ecen}{\end{center}}

\newcommand{\eps}{\varepsilon}
\newcommand{\lam}{\lambda}

\newcommand{\f}[2]{\frac{#1}{#2}}
\renewcommand{\b}[1]{\left({#1}\right)}

\renewcommand{\sb}[1]{\left[{#1}\right]}
\newcommand{\mean}[1]{\langle {#1} \rangle}

\usepackage{xcolor}

\newcounter{lastnote}

\title{Controlling the uncontrollable: Quantum control of open system dynamics}

\author
{Shimshon Kallush,$^{1,2}$ Roie Dann,$^{2}$ Ronnie Kosloff$^{2}$\\
\\
\normalsize{$^{1}$Sciences Department, Holon Academic Institute of Technology,}\\
\normalsize{52 Golomb Street, Holon 58102, Israel}\\
\normalsize{$^{2}$The Institute of Chemistry, The Hebrew University of Jerusalem, Jerusalem 9190401, Israel}\\
\\
\normalsize{$^\ast$Ronnie Kosloff; E-mail:  ronnie@fh.huji.ac.il}
}

\date{}

\begin{document} 

% Double-space the manuscript.

\baselineskip24pt

% Make the title.

\maketitle

% Place your abstract within the special {sciabstract} environment.

\begin{sciabstract}
Control of open quantum systems is an essential ingredient to the realization of contemporary quantum science and technology. We demonstrate such control by employing a thermodynamically consistent framework, taking into account the fact that the drive can modify the interaction with environment. 
Such an effect is incorporated within the dynamical equation,
leading to control dependent dissipation, this relation serves as the key element for open system control. Thermodynamics of the control process is reflected by a unidirectional flow of energy to the environment
resulting in large entropy production.
The control paradigm is displayed by analyzing 
entropy changing state to state transformations, such as heating and cooling.
In addition, the generation of quantum gates under dissipative  conditions is demonstrated for both non-unitary reset maps with complete memory loss
and a universal set of single and double qubit unitary gates.
\end{sciabstract}

% In setting up this template for *Science* papers, we've used both
% the \section* command and the \paragraph* command for topical
% divisions.  Which you use will of course depend on the type of paper
% you're writing.  Review Articles tend to have displayed headings, for
% which \section* is more appropriate; Research Articles, when they have
% formal topical divisions at all, tend to signal them with bold text
% that runs into the paragraph, for which \paragraph* is the right
% choice.  Either way, use the asterisk (*) modifier, as shown, to
% suppress numbering.

\section*{Introduction}
Quantum control  addresses the task of driving the state of a quantum system to a desired objective. This is achieved by applying coherent control fields which orchestrate the interference of quantum amplitudes, i.e., quantum coherence \cite{rice1992new,warren1993coherent}. Coherent control has been successfully applied for a variety of tasks \cite{glaser2015training}, however, the key ingredient, coherence, remains extremely sensitive to any external perturbation. Realistically, all quantum systems are to some extent open, thus, are
subject to environmental effects. Interaction between the device and the external environment generates system-environment correlations. These in turn  effectively degrades the required agent, coherence, leading to a detrimental effect on coherent control \cite{wu2007controllability,koch2016controlling,schmidt2011optimal,lloyd2000control}.  Nevertheless, the inevitable ``harmful'' dissipation also allows to redefine the possible control objectives by enabling non-unitary
entropy changing transformations
\cite{dann2019shortcut,dann2020fast}. 

The present study explores such entropy changing control targets. The basic proposition is based on the realization that the external drive influences not only the primary system (directly) but also the dissipation induced by the environment (indirectly). We will demonstrate how the interplay between direct and indirect
control can lead to the important building blocks of quantum coherent control, such as, unitary gates under dissipative conditions and irreversible reset operations. In addition, we analyze the thermodynamic consequences of the open system control processes. Rapid control protocols are found to require additional heat dissipation to the environment and the
unitary gates are accompanied by active cooling in order to maintain high purity of the systems, with the price in large entropy production.

The control relies on 
the intimate relation between the isolated system (free) dynamics and the dissipative part of the dynamics. This relation is a consequence of a global Hamiltonian $\hat{H}$, which includes a quantum description of the system, controller and environment.
%Specifically, we employ time-translation symmetry to obtain the equations of motion required for control. As desired, these equations comply with thermodynamic principles \cite{dann2020thermodynamically}. 
Our control objective is defined solely in terms of system observables, which are subject to the environmental influence.

Such control process is described within the framework of  open quantum systems \cite{breuer2002theory}, where the reduced description is given by an appropriate non-unitary dynamical equation of motion. %\tg{Description of such a process requires the framework of the theory of open quantum systems. Here, the key element is the reduced description by an appropriate dynamical equation of motion.}  
Assuming negligible initial correlations between the system and environment, the reduced dynamics are governed by a  completely positive trace preserving map (CPTP)
$\hat \rho_S (t)=\Lambda_t \hat \rho_S(0)$
\cite{kraus71}. This map is generated by the dynamical equation
\begin{equation}
    \frac{d}{dt} \hat \rho_S( t)={\cal L}_t \sb{\hat \rho_S( t)}~~.
\end{equation}
The precise form of the generator is obtained by  a first principle `microscopic' derivation (Appendix \ref{appendix-A}). 
The equation is valid under weak coupling  between the system and environment and a timescale separation between the slow system and fast environment. Thermodynamically this dynamics represents an isothermal partition, allowing heat
transfer leading to thermal equilibrium \cite{dann2021open,dann2020thermodynamically}.

 The  control dynamical equation is of the  Gorini Kossakowski Lindblad Sudarshan (GKLS)  form \cite{gorini1976completely,lindblad1976generators}
\begin{equation}
  \frac{d}{dt}{{\hat\rho_S(t)} }=   -\frac{i}{\hbar}[{\hat{H}_S( t )},{ \hat{\rho}_S (t)}] +\mathcal{L}_d \sb{\hat{\rho}_S( t)}~~,
  \label{eq:lindblad}
\end{equation}
where the dissipative part 
${\cal L}_d$ has the  structure 
 \begin{equation}
     {\cal L}_d\sb{\hat \rho_S}=\sum_{j}\gamma_{j}(t)
    \bigg( \hat{F}_{j}( t) \hat{\rho}_{S}( t)\hat{F}_{j}^{\dagger} ( t )\\-\frac{1}{2}\{\hat{F}_{j}^{\dagger}( t)\hat{F}_{j} ( t),\hat{\rho}_{S}( t) \}\bigg) ~~.
    \label{eq:gen_NAME}
 \end{equation}
Here,  the Lindblad jump operators $\hat F_j$ constitute eigenoperators of the free dynamical map ${\cal U}_{S}(t)$, Eq. (\ref{eq:eigenf}), and the kinetic coefficients $\gamma_{j}(t)$  are real and positive. 
%In order to simplify the exposition we neglected in Eq. (\ref{eq:lindblad}) the Lamb-shift and the pure dephasing terms. For further details regarding the derivation of the master equation see Appendix \ref{appendix-A}.
 %\tb{maybe say something about the fact that they can be also negative, or assume semi-group property (strict Markovianity) and then they must be positive.}.
 
 Turning off the controller
 $\hat V(t)=0$ the system will settle to thermal equilibrium with the drift Hamiltonian $\hat H_S^0$.
 The time dependent controller $\hat V(t)$ modifies
 the fixed point of the equation: ${\cal L}_t\sb{\hat \rho_S^{i.a.}}=0$ to which the system aspires; termed the instantaneous attractor
 .
 
The  controller influences the system state both  directly, through the unitary term, and indirectly, through the jump operators and kinetic coefficients of the dissipative part.
 
The dual structure suggests an iterative approach to extract the control field: 
\begin{enumerate}[label=(\roman*)]
\item
Guess a control field and apply it to calculate an explicit solution of the system's free dynamics ${\cal U}_S(t)$. 
\item
Construct the master equation according to Eq. (\ref{eq:lindblad}).
\item
Calculate the evolution, $\hat{\rho}_S (t)$.
\item
Utilizing the final state evaluate
the control objective. 
\item
Using the evaluated control objective functional, update the field and iterate the cycle until convergence. 
\end{enumerate}

In step (i)
we construct ${\cal U}_S (t)$ from the unitary evolution operator: ${\cal U}_S(t,0) [{\bullet}]= \hat U_S(t,0) \bullet \hat U^\dagger_S(t,0)$,
generated  by the ``semi-classical Hamiltonian'' $\hat H_S(t)$:
\begin{equation}
    i\hbar \frac{\partial}{\partial t} \hat U_S (t) = \hat H_S (t) \hat U_S(t)~~,
    \label{eq:evol}
\end{equation}
with $\hat U_S (0)=\hat I$. Here, the semi-classical Hamiltonian is composed of the bare-system Hamiltonian and control term 
\begin{equation}
    \hat{H}_S ( t )= \hat{H}_S^0 +\hat{V}( t)~~,
\end{equation}
where $\hat{V}(t)$ is a system time dependent control operator. It can be derived from the autonomous description of the control state and system-control interaction, cf. Ref \cite{dann2020thermodynamically} for further details.
For specific control protocols Eq. (\ref{eq:evol}) posses closed form solutions, which can be extended for slow deviations from such protocols utilizing the inertial theorem \cite{dann2021inertial}. But a general analytical solution requires overcoming a time-ordering procedure \cite{dyson1949s}.

In this study, we bypass the time-ordering obstacle by employing a
numerical solution of the free dynamics Eq. (\ref{eq:evol}).
This leads to the eigenstates of the time-evolution operator
\begin{equation}
    \hat U_S(t) |\phi_n(t)\rangle= e^{-i\epsilon_n \b t}|\phi_n \b t\rangle~~.
\end{equation}
From $|\phi_n\b t\rangle$ we construct the eigenoperators
\begin{equation}
    \hat F_j (t) = | \phi_n \b t \rangle \langle \phi_m (t) |~~,
    \label{eq:jump}
\end{equation}
 where $j=N(n-1)+m$.
These satisfy the eigenvalue-type relation with respect to the free propagator
\begin{equation}
{\cal U}_S(t,0) \hat F_j = \hat U_S(t,0) \hat F_j\hat U^{\dagger}_S(t,0) = e^{-i\theta_j\b t} F_j~~,
    \label{eq:eigenf}
\end{equation}
where $\theta_j\b t=\eps_n\b t-\eps_m \b t$ are the corresponding phases. They determine the effective instantaneous Bohr frequencies of the system $\omega_j\b t=d\theta_j\b t/dt $.
The non-invariant eigenoperators come in conjugate pairs with complex conjugate eigenvlaues, and
%\tg{$\hat F_j^{\dagger}$ with eigenvalue $\lambda_j (t) =\exp(i\omega_j(t) t) $.}
constitutes transition operators between the instantaneous eigenstates of $\hat{U}_S ( t)$. While, eignoperators with $\theta_j \b t=0$ are  the instantaneous projection operators, $\{ | \phi_m \rangle \langle \phi_m |\}$ .

The remaining task to obtain the control dynamical equation (\ref{eq:lindblad}) (step (ii)) is to calculate the kinetic coefficients $\{ \gamma_j\b{t}\}$. The fact that the jump operators associated with a certain transition are related by a detailed balance relation motivates relabeling the kinetic coefficients  $k_{i,\uparrow}\b t ~,~k_{i,\downarrow} \b t$, where $i=j/2$ corresponds to a conjugate pair of eigenopertors.
In the weak coupling limit and under Markovian dynamics, these coefficients can be calculated from the Fourier transform of the environment correlation functions with instantaneous frequency $\omega_j\b t$ \cite{dann2018time,dann2021non}. The dynamical equation leads to $\hat{\rho}_S (t)$ (step iii), which allows calculating the control objective. Finally, the objective is used to update the control field (steps (iv) and (v)).

\section{Model}
\label{sec:model}

We demonstrate the quantum control scheme by studying open system control scenarios of the single mode Bose-Hubbard model \cite{vardi}. For  $N$ particles in a double well potential, the system is isomorphic to the angular momentum Hamiltonian with $j=N+1$, where $j$ is the total angular momentum
\begin{equation}
    \hat{ H}_S^0 =  u\hat{ J}^2_z +\Delta\hat{ J}_x~~. 
    \label{eq:hstat}
\end{equation}
Here $\hat{ J}_x$ is the hopping operator and $\hat{J}_z^2 $ is the in-site interaction operator. We set $u = 2\Delta/j$ for which the dynamics are classically chaotic \cite{trimborn2008exact}, and employ the control Hamiltonian
\begin{equation}
    \hat{ V}(t) = \epsilon(t)\hat{ J}_z~~,
    \label{eq:hcont}
\end{equation}
where $\hat{ J}_z$ controls the energy balance. 

We chose this model since the free dynamics of the system, including both Hamiltonians Eqs. (\ref{eq:hstat}) and (\ref{eq:hcont}) is completely unitary controllable \cite{huang1983controllability,kallush2011scaling}. Moreover, the same control Hamiltonian is scalable  to an arbitrary $N$-level system. 

The dynamics of the closed system evolution operator $\hat{  U}_S ( t)$, Eq. (\ref{eq:evol}) is integrated numerically by a Chebychev propagator \cite{tal1984accurate}. At each intermediate time  $ \hat{ U}_S (t)$ is  diagonalized to obtain the time-dependent orthonormal set of jump operators $ \{\hat{ F}_j( t)\}$, Eq. (\ref{eq:jump}), and Bohr frequencies $\{\omega_j(t)\}$. The jump operators are then employed to compute the Liouvillian dynamics in the interaction picture. Employing the Liouvillian superoperator, the full dissipative equation of motion, Eq. (\ref{eq:lindblad}),
was propagated numerically for $\hat \rho_S\b t$ employing a Newtonian polynomial method \cite{ashkenazi1995newtonian,kosloff1994propagation}.

The environment was chosen as a bosonic Ohmic bath with a spectral density $J(\omega)= c\omega^2 $, where $c$ is scaling constant which takes care of units. Such a choice corresponds to an interaction with the electromagnetic field or a phonon bath. %\tg{The interaction with the environment was explicitly constructed to comply with the global symmetry constraints.}
For each term in the sum of Eq. (\ref{eq:gen_NAME}) the corresponding kinetic coefficients are functions of the Bohr frequency $\omega_j$
\begin{eqnarray}
k_{j,\uparrow} ( t)=g^2\omega_j( t) J(\omega_j)N(\omega_j( t))=k_{j,\downarrow}(t) e^{-\hbar \omega_j( t)/k_B T}~~,
\label{eq:kinetic_coeff}
\end{eqnarray}
where $N(\omega)=1/(e^{\hbar \omega/k_BT} -1)$, and
$g$ is the system-environment coupling. For the numerical analysis we set the parameters such that  $g^2c = 10^4$ in atomic units and system-bath coupling operator is taken to be proportional to $\hat{J}_y$. 

%The field free dynamics under stationary dissipation were verified to lead to the expected thermal states. 

The control scheme that was employed to evaluate the optimal field is a simplified version of the CRAB algorithm \cite{caneva2011chopped,rach2015dressing}. For each task a cost function was defined (see below) and the field is given by
\begin{equation}
    \epsilon (t) = \exp\b{ -\b{{\frac{t-\tau/2}{2\sigma}}}^2} \sum_{k=1}^M c_k \sin(\nu_k t) 
    \label{eq:cont_exp}
\end{equation}
where $\sigma$ is the pulse width, $\tau$ the target control time and $\nu_k$ are a set of $M$ frequencies. 
The coefficients $c_k$ where varied to optimize the cost function, utilizing a standard quasi-Newton algorithm. In the present study the amplitude of the control field was not constrained. Nevertheless, one can include additional constraints within this CRAB-like method, such as a total pulse energy restriction of the form $\lambda\int \epsilon(t)^2dt$ or entropy generation. 

\section{Control Results}
\label{sec:results}

To illustrate the control scheme, we first demonstrate state-to-state entropy changing tasks and proceed by analysing control of dynamical maps.
In all the cases studied the control landscape was found to contain traps, meaning that  sub-optimal minima exist. We overcome this difficulty by employing hundreds of realizations with different random initial guesses for the field. The solution presented are the best from this set.

\subsection{Heating and Cooling}
\label{subsec:heating-cooling}

The hallmark of open system control 
is a change in the system von Neumann entropy
\begin{equation}
    {\cal S} = -\rm{tr}\b{\hat \rho_S \log ({\hat \rho_S})}~~.
    \label{eq:vnentropy}
\end{equation}
The requirement is a clear indication of an interaction with an external environment since unitary control necessarily preserves the eigenvalues of 
$\hat \rho_S$, ${\cal S}$ must be constant for isolated systems.
This property motivates the choice of  the entropy ${\cal S}$ as our cost functional for the state-to-state control
objective.
Heating or cooling are defined by
 an increase or decrease of  the systems von Neumann entropy, correspondingly.
For the demonstration we choose a thermal initial  state  $\hat{\rho}^{i}_S$ with temperature $\beta \equiv {\frac{1}{k_B T}} = {\frac{1}{\Delta}}$. 

During an the open system dynamics the system and the environment entropies vary. The total amount of entropy produced can be evaluated by integrating the entropy production rate:
\begin{equation}
    \Sigma^U\b t\equiv -\f{d}{dt}{\cal D}\b{\hat{\rho}_S |\hat{\rho}_S^{i.a}}=-k_{B}\rm{tr}\b{{{\cal L}}_t \sb{\hat \rho_S}\log \hat{\rho_S}}+k_{B}\rm{tr}\b{{\cal L}_t \sb{\hat \rho_S}{\log} \hat{\rho}_S^{i.a}} 
    \label{eq:entropy_production_name}
\end{equation}
where ${\cal D}$ is the divergence and $\hat \rho_S^{i.a}$ is the time dependent instantaneous attractor \cite{dann2019shortcut}, which satisfies ${\cal L}\sb{\hat{\rho}_S^{i.a}}=0$.
By integrating over Eq. (\ref{eq:entropy_production_name}) over the protocol duration one obtains the total entropy production.

{\bf Heating}: Our current control task is to heat the system as much as possible. For an $N$ level system, this task defines the target state as the micro-canonical distribution $\hat{\rho}_S^f = \hat{I}/N$, with the maximal entropy ${\cal S}^{\rm{max}} = \log N$.

\begin{figure}
\centering
\includegraphics[width=8.7cm]{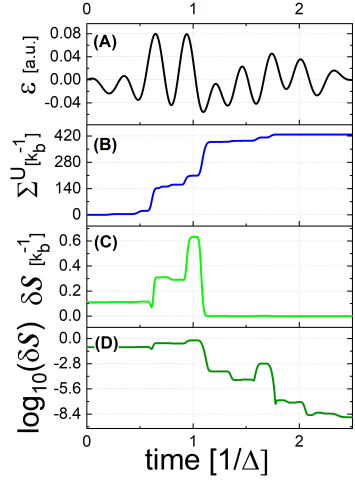}
    \caption{
    Control process of the Two Level System Heating. Time defined in units the inverse frequency 
    $\Delta$ Eq. (\ref{eq:hstat}): {\bf (A)} The optimized control field. {\bf (B)} Accumulated entropy production into the environment, obtained by the time integration of 
    the entropy production rate Eq. (\ref{eq:entropy_production_name}).
   {\bf (C)} Divergence of the system's entropy from its maximal value in linear scale. {\bf (D)} Similar to {\bf (C)} in logarithmic scale. }
    \label{fig:TLS}
\end{figure}

Figure \ref{fig:TLS} demonstrates a controlled heating task for a two level system (TLS). We find that initially coherence is generated and the system entropy decreases. While at the final stage, the dissipation of coherence is accompanied by significant heating, leading  
to an entropy production of about three orders of magnitudes larger then system's change in entropy. This result can be understood by the fact that the optimization was performed only with respect to system entropy, while the dissipative entropy generation was not constrained. 

The protocol was calculated by performing an optimization over the control space, which corresponded to  $M=20$ field frequencies, Eq. (\ref{eq:cont_exp}). In addition, the timescale of the control pulse is chosen to be inversely related to the TLS energy difference $2 \pi/\Delta$, which is much shorter compared to the chosen natural spontaneous decay rate, given by ${10^{-4}}/{\Delta}$. 
This boost in performance stems from the dependence of the kinetic coefficients $\{\gamma_j\}$, Eq. (\ref{eq:gen_NAME}), on the driving parameters. The indirect control over the kinetic coefficients leads the maximal entropy state with a precision of $10^{-9}$.

\begin{figure}
    \centering
    \includegraphics[width=9cm]{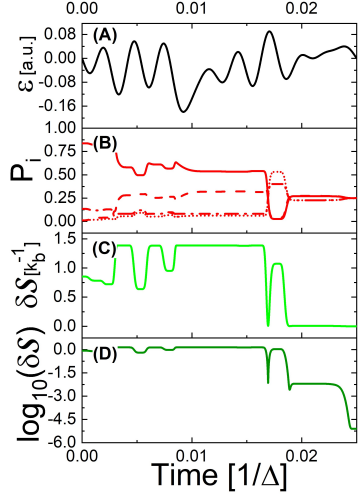}
    \caption{Control of four-level system; Heating ($N=4$).  Similar to fig. \ref{fig:TLS}, except for  \textbf{(B)} which presents the population of the system's four states as a function of time. At the final time a uniform distribution is obtained, which corresponds to an infinite temperature thermal state.}
    \label{fig:FLS}
\end{figure}

The same maximum entropy objective has been employed for four levels Cf. Fig. \ref{fig:FLS}. The target fidelity  reached 
a relative error of  $10^{-5}$ compared to the target  maximal entropy $S^{\rm{max}} = \rm{ln}(N)$.
This is significant reduction with respect to the qubit case.     The optimal control duration for the four level system was had comparably a shorter protocol duration,  Cf. Fig. \ref{fig:FLS}.
The complexity of the control increases with
the number of possible interference paths, and  grows exponentially with the number of levels and control duration. 
This fact influences the optimization procedure, finding the optimal control objective becomes more demanding with
the increase in size. 
As a result the target fidelity is reduced as well as the effective time window
a solution can be found.

\begin{figure}
    \centering
    \includegraphics[width=9cm]{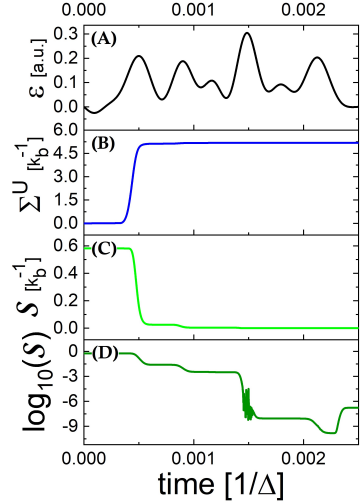}
    \caption{ Control of Two-levels system cooling: Designation similar to Fig. \ref{fig:TLS}.  The  target of control the system's entropy is displayed in two panels \textbf{(C)} and \textbf{(D)}  in linear and logarithmic scales.} 
    \label{fig:coolTLS}
\end{figure}

{\bf Cooling}: From a formal algorithmic point of view, cooling is almost identical to heating, but with a modified objective. Here, the goal is to minimize the system's entropy, therefore, the final
target state is pure, satisfying $({\hat \rho_S^f})^2=\hat \rho_S^f$. Moreover, in both processes
once reaching the target state, where the coherence vanishes, the expectation value of the controller vanishes resulting in the system becoming  strictly uncontrollable. 
Physically, however, the two processes significantly differ from one another. 
The asymmetry has its origins within the third law of thermodynamics, which implies that the resources required to cool to zero entropy diverge \cite{levy2012quantum,masanes2017general,taranto2021landauer}.    
 Fig. \ref{fig:coolTLS}, presents the controlled cooling of a TLS. The objective, which is the minimization of the final state entropy,
is obtained with high accuracy. However, we find that achieving extremely low entropy values typically require large amplitude of the control fields, leading to numerical instabilities.
A possible remedy
is to introduce an extra term in the cost function, preventing the abusive use of resources by the control. 

\begin{figure}
    \centering
    \includegraphics[width=8.5cm, height=9cm]{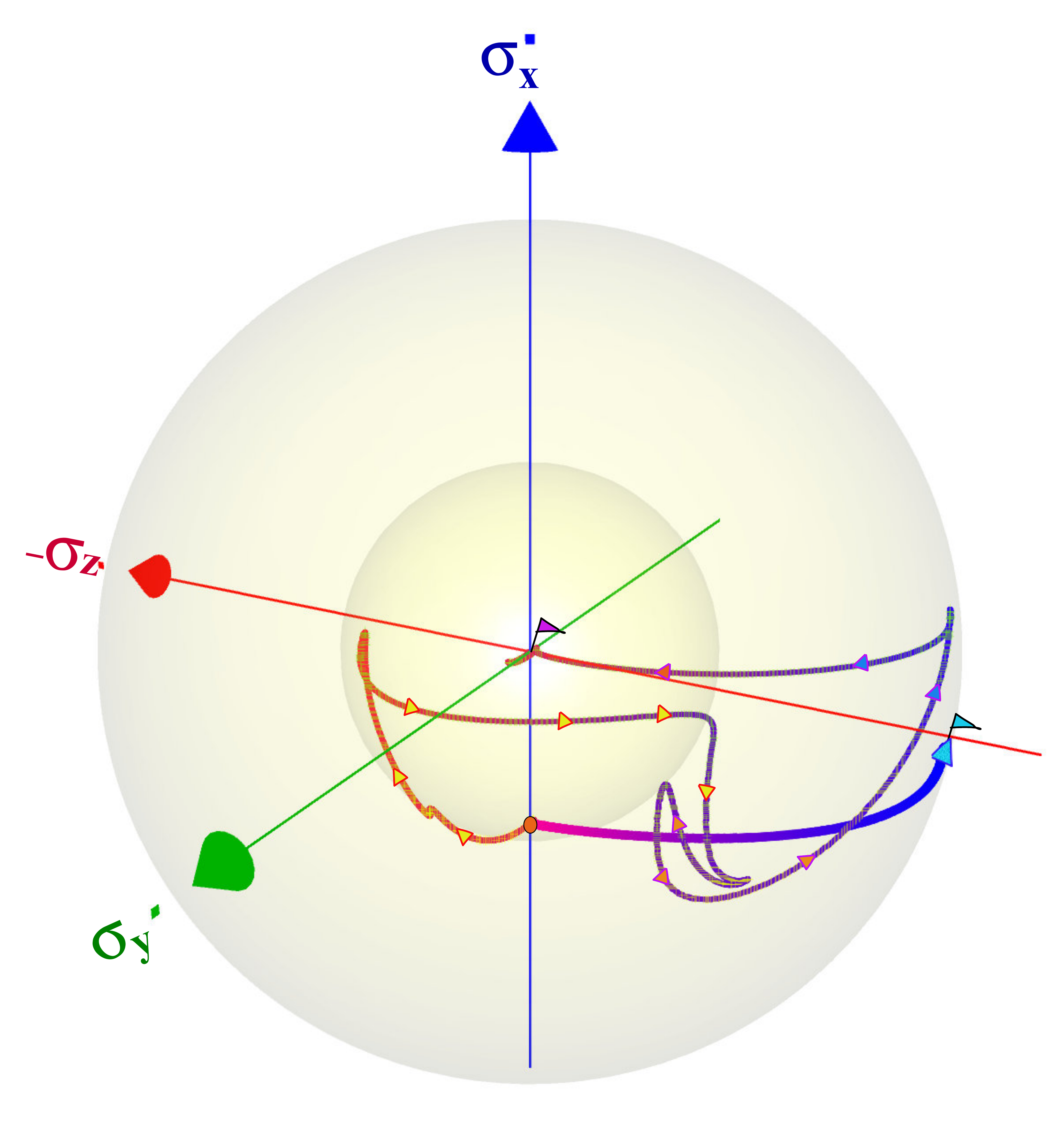}
    \caption{The control trajectories of the heating and cooling solutions displayed on the Bloch sphere.
    The common initial thermal state is designated by an orange dot on the $x$ axis.
    The cooling trajectory monotonically approaches a pure state on the surface of the Bloch sphere in the $-\sigma_z$ direction, while the heating trajectory first increases purity (The inner sphere represents the initial purity). In the final stage, purity decreases as the trajectory approaches the completely mixed state at the origin.} 
    \label{fig:bloch-ball}
\end{figure}

A control trajectory in the Bloch sphere, for optimal heating and cooling  processes, is shown in Fig. \ref{fig:bloch-ball}. The projection of the instantaneous $\hat \rho_S(t)$ on the three Pauli operators is shown  on the Bloch sphere.
In this geometrical representation of the TLS, the state's distance from the origin represents its purity, and any state which lacks coherence, such as the thermal state, resides on the $x$ axis The cooling and heating trajectories initialise at the same thermal state, designated by an orange dot. Heating takes the initial state to the origin, via a trajectory that passes through the high radii region, corresponding to states with high purity.  Cooling is achieved by a more direct path to an almost final pure state. Comparing the control protocols, the cooling protocol requires higher instantaneous power compared to the heating process. This results in an increase in the effective Rabi-frequency for the cooling, which agrees with the overshoot observed in the shortcut to equilibration protocol \cite{dann2019shortcut,dann2020fast}.

\subsection{Control of dynamical maps}
\label{subsec:maps}

Generation of a  CPTP dynamical map \cite{kraus1971general}:  
$\hat \rho_S^f =\Lambda \hat \rho_S^i$ constitutes a more stringent control task. A map must transform any arbitrary initial state to the corresponding target state. Compared to an level
state to state transformation a unitary gate is expected to be facrorially more difficult
to achieve.
The map transformation can be fully characterized by
employing a complete operator basis and employing the scalar product:
$(\hat A, \hat B)=\rm{Tr} \{
\hat A \hat B^{\dagger}\}$. For example, in the qubit case,
we can express the state using the set of Pauli operators and the identity, $\{ \hat I,\hat \sigma_x,\hat \sigma_y,\hat\sigma_z\}$.  The map $\Lambda$ 
can therefore be expressed in terms of a $4 \times 4$ matrix.

Two extreme cases are studied:
A reset map $\Lambda_R$ and unitary map $\Lambda_U$.
The reset map transforms any initial state to a single
target state. Specifically, considering an arbitrary initial state
 \begin{equation}
    { \hat{\rho}}_{S}^{i} = {\frac{1}{2}} { \hat{I}} + \sum_{j=x,y,z} c_j { \hat{\sigma}}_j~~,   
\end{equation}
the chosen map transforms any state to a pure state in the $x$ direction
\begin{equation}
    { \hat{\rho}}^f_S =  {\frac{1}{2}}
\left(\begin{array}{rr}
1 & -1 \\
-1 & 1 
\end{array}\right)
= {\frac{1}{2}} ({ \hat{I}}- { \hat{\sigma}}_x)~~.
\label{eq:tarent}
\end{equation}
In the operator space, spanned by $\{\hat{I},\hat{\sigma}_x,\hat{\sigma}_y,\hat{\sigma}_z\}$ the associated transformation is represented by the non-unitary matrix 
\begin{equation}
\Lambda_R = 
\left(
\begin{array}{ccccc}
1 & 0 & 0 & 0\\
0 & -1 & -1 &-1 \\
0 & 0 & 0 & 0 \\
0 & 0 & 0 & 0 
\end{array}~~.
\right)
\end{equation}

\begin{figure}
    \centering
    \includegraphics[width=9cm]{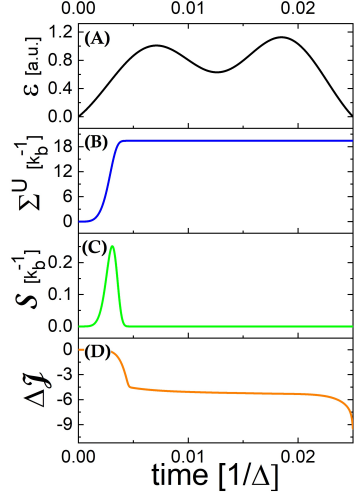}
    \caption{Control of reset transformation: Transformation of an arbitrary qubit state to a final pure state $\hat{\rho}^f_S$, Eq. (\ref{eq:tarent}). {\bf{(A)}} Time dependent control field. {\bf{(B)}} Environmental entropy production. {\bf{(C)}} System's entropy.  {\bf{(D)}} Transformation infidelity in logarithm scale, as a function of time.}
    \label{fig:reset}
\end{figure}

To determine the generating field $\epsilon(t) $, Eq. (\ref{eq:hcont}), a complete set of initial states ~$\left\{ \hat \rho_S^{i,k} \right\}$ is employed and optimized to reach the same target state $\hat \rho_S^f$. The accuracy of the transformation is then evaluated by the objective functional, the trace distance 
\begin{equation}
    {\cal{J}} =  \sum_{k}{\rm{tr}}\{\hat \rho^{f,k}_S \hat \rho^f_S\}~~,
    \label{eq:objective}
\end{equation}
where $\hat{\rho}_S^{f,k} = \Lambda \hat{\rho}_S^{i,k}$.
Figure \ref{fig:reset} demonstrates the  reset transformation.  As shown in Panel {\bf{(C)}}, at initial and final times  the systems are in a pure state, while at intermediate times an increase in entropy indicates the necessary temporary transition to a mixed state.
The obtained mechanism of reset process can be divided into two stages: At the beginning we witness an entropy increase, indicating a loss of memory of the initial state. This is followed by a purification of the mixed state rotation to the desired direction, in the final stage.

Panel {\bf{(D)}} presents the deviation of the system's objective functional from its maximal value in a logarithm scale. An initial rapid reduction in the deviation brings the system to a precision of ${\Delta {\cal J}}=10^{-5}$. This is followed by a final stage giving an additional accurate kick, which drives the system to the target state and to deviations of up to $10^{-9}$. Crucially, we also verified explicitly that the obtained control field transforms any randomly picked pure and non-pure state into the target state, which is, indeed the manifestation of the reset transformation.  Note that the meaning of such reset transformation constitutes an ultimate cooling process of the  system. That is, the obtained field cools the system effectively from any initial state to $T\approx 0$. As expected, since no restriction was imposed on the entropy production rate, the thermodynamic cost of the reset process, exhibited in panel {\bf{(B)}}, is well above its theoretical bound given by the Landauer limit \cite{landauer1961irreversibility}. 

\subsection{Unitary maps}
\label{subsec:i-maps}

We next tackle the task of inducing a unitary transformation under dissipation
\cite{kallush2006quantum,goerz2014optimal,wenin2008state}. 
The chosen demonstrative transformations are the one qubit Hadamard $\Lambda_U$ and a two qubit entangling gates $\Lambda_{\sqrt S}$. Since an entangling gate and the full set of single qubit rotations form a universal set of quantum gates, by combining such control protocols an arbitrary unitary gate can be achieved    \cite{nielsen2002quantum,muller2011optimizing,watts2015optimizing,goerz2015optimizing}. 
These transformations can be incorporated in noisy quantum information processing, producing effective unitary single qubit and two qubit gates under dissipation. 

\begin{figure}
    \centering
    \includegraphics[width=14cm]{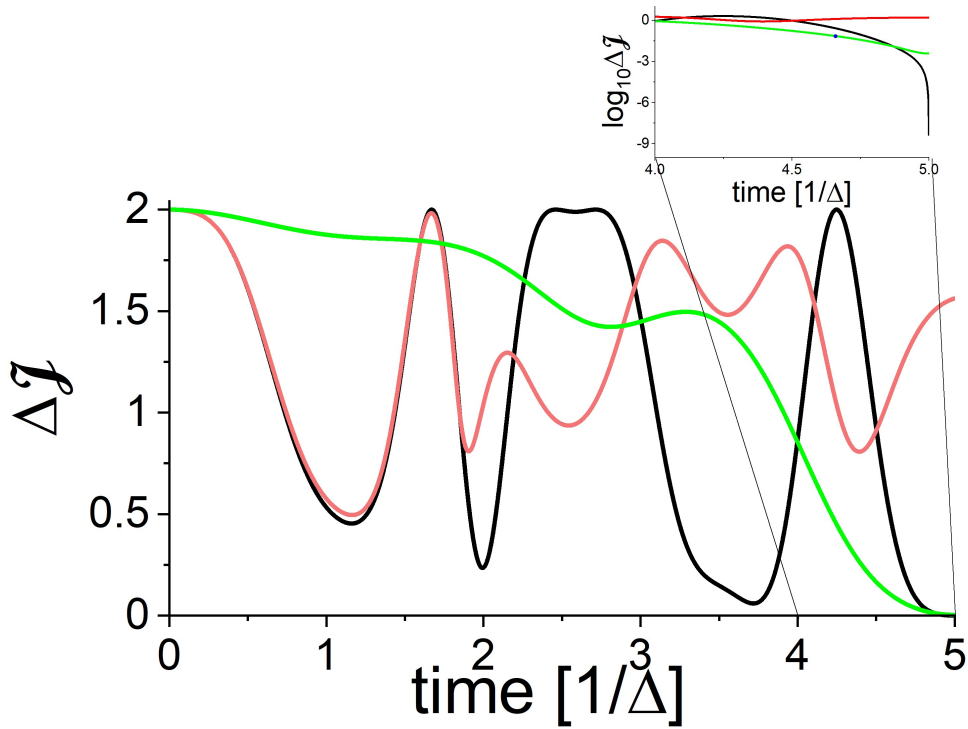}
    \caption{ Hadamard gate control: Deviation of the objective functional, Eq. (\ref{eq:objective}), as a function of time for the control leading to the Hadamard gate, Eq. (\ref{hadtran}).  (main) Linear and (inset) logarithm scales. (black) Optimal transformation under dissipation-free propagation. (red) The dynamics of the transformation with the same field subject to the environment (relaxation time of $\tau = 10^{-11}\rm{sec}$). (green) Optimal dynamics for the open system. }
    \label{fig:unit}
\end{figure}

\begin{figure}
    \centering
    \includegraphics[width=12cm]{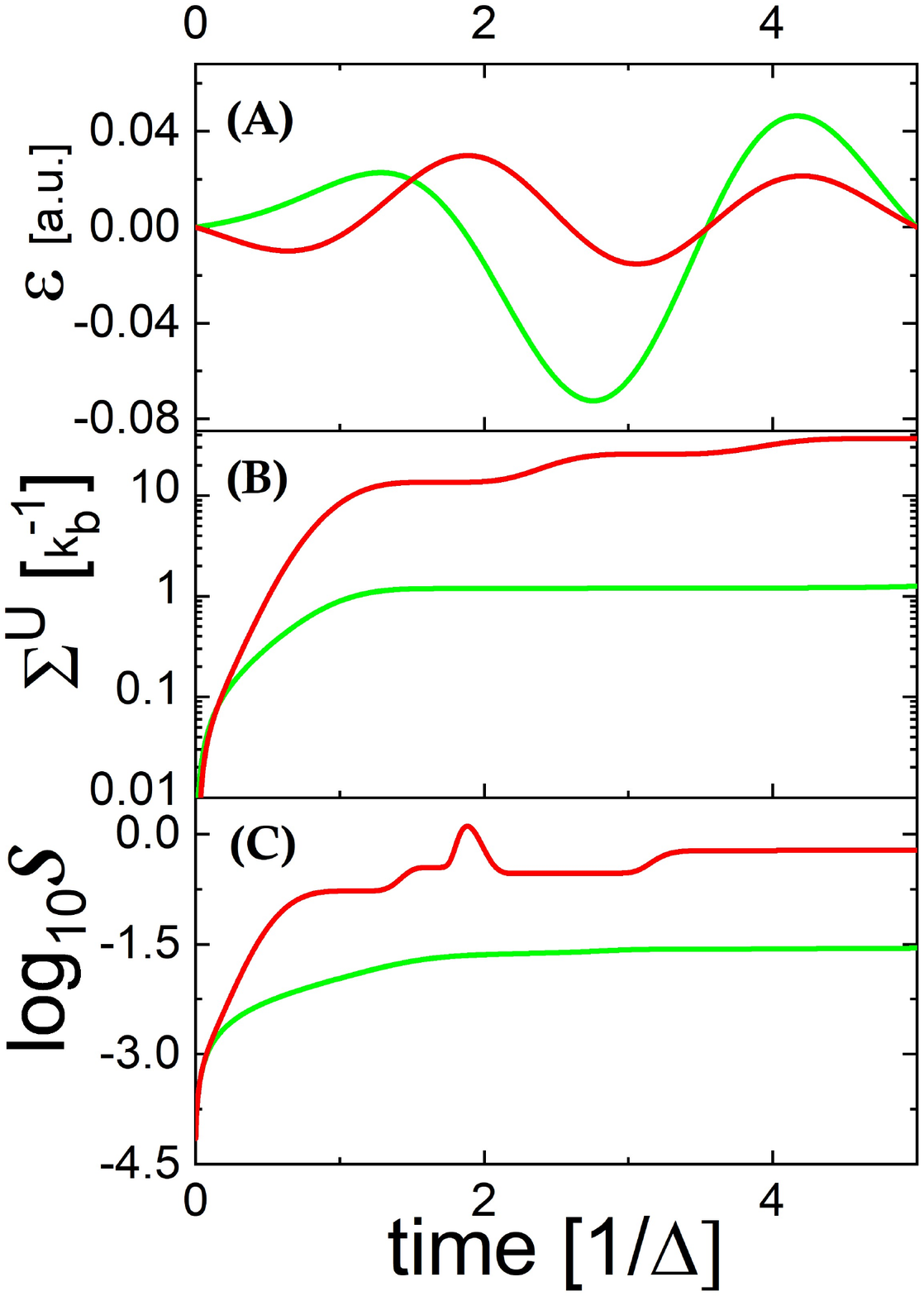}
    \caption{ Hadamard control: {\bf(A)} Control field, {\bf (B)} entropy productio and {\bf (C)} system's entropy as a function of time for the Hadamard transformation. Color coding is similar to Fig. \ref{fig:unit}. The field for the open system dynamics is increased by a factor of ten to enable comparison. }
    \label{fig:unit1}
\end{figure}

%\textbf{???Why do we need this? ???}
%The general single qubit gate is equivalent to
%the rotation map:
%\begin{equation}
%\Lambda_O = 
%\left(
%\begin{array}{cc}
%1 & 0 \\
%0 & {\cal R}_{\chi}(\theta)
%\end{array}
%\right)~~,
%\label{eq:rotations_single}
%\end{equation}
%where ${\cal R}_{\chi}(\theta)$ is the $3 \times 3$  rotation matrix around axis $\chi$ at an angle $\theta$.
%Alternatively, the map can be completely characterized  in terms of three eigenoperators:
%$\hat{\sigma}_{\chi}$ and  $\hat{\sigma}_{\chi +}, \hat{\sigma}_{\chi-}$ with corresponding eigenvalues $1,e^{i\theta}$ and $e^{-i \theta}$. As a consequence, these  three operators are sufficient to determine the accuracy of the map. 
%\textbf{Redundant:}
%\textbf{We next demonstrate the single qubit control by finding a protocol which generates the Hadamard gate,}
A single qubit gate corresponds to a rotation in the Bloch sphere and can be expressed as a superoperator in $\{\hat{I},\hat{\sigma}_x,\hat{\sigma}_y,\hat{\sigma}_z\}$ the operator basis. Specifically, the Hadamard gate is given by
\begin{equation}
\Lambda_U = 
\left(
\begin{array}{cccc}
1 & 0 & 0 & 0\\
0 & 0 & 0 & -1 \\
0 & 0 & -1 & 0 \\
0 & -1 & 0 & 0 
\end{array}
\right)~~.
\label{hadtran}
\end{equation}
The algorithm leading to the optimal field is similar to the one described in subsection \ref{subsec:maps}. Namely, we initialize the system with a complete set of pure density operators  $\left\{ \hat \rho_{S}^{i,k} \right\}$ and chose the cost function as the sum of trace overlaps between the final and target states. The results are presented in Figs. \ref{fig:unit} and \ref{fig:unit1}.
To get a measure for the validity and robustness of the protocol we studied three different scenarios:
\begin{enumerate}[label=(\alph*)]
    \item{First, the optimization was carried for an isolated system. Such a protocol coincides with the conventional closed system unitary transformation. The results of this optimization are presented in the black curves in the various panels of the figures. One can see that the transformation is achieved with the expected high accuracy.}
    \item{The same optimal field of (a) was applied to the open quantum system. The results are shown in red. As can be seen, the dynamics  at earlier times seem similar, but deviate significantly at later times. The final  precision $\Delta{\cal J}$,  is well above the reliable operational threshold of  feasible gates. As can be seen in Fig. \ref{fig:unit1} the degradation in precision is accompanied by an undesired increase in entropy, which stems from the  coupling to the environment.}
    \item{Finally, the optimization was generated from scratch, taking into account the full open system dynamics. The associated results are presented by green curves. Accounting for the external dissipation allows the control to cope with the environmental noise. Despite of the strong decoherence, the unitary transformation  precision is below the threshold of $10^{-3}$, well within the acceptable specs of feasible quantum gates. The presence of a relatively strong system-environment coupling leads to generation of entropy. Nevertheless, it is reduced with respect to the reference protocol and the entropy leak is suppressed by the field at later times.}
\end{enumerate} 
Surprisingly, we observe that required control field amplitude for the open system dynamics is significantly lower than the free dynamics control field (see right panel of Fig. \ref{fig:unit1}). As a consequence, the total energy employed by the optimal field is smaller by two orders of magnitude.

\begin{figure}
    \centering
    \includegraphics[width=8.5cm, height=8.5cm]{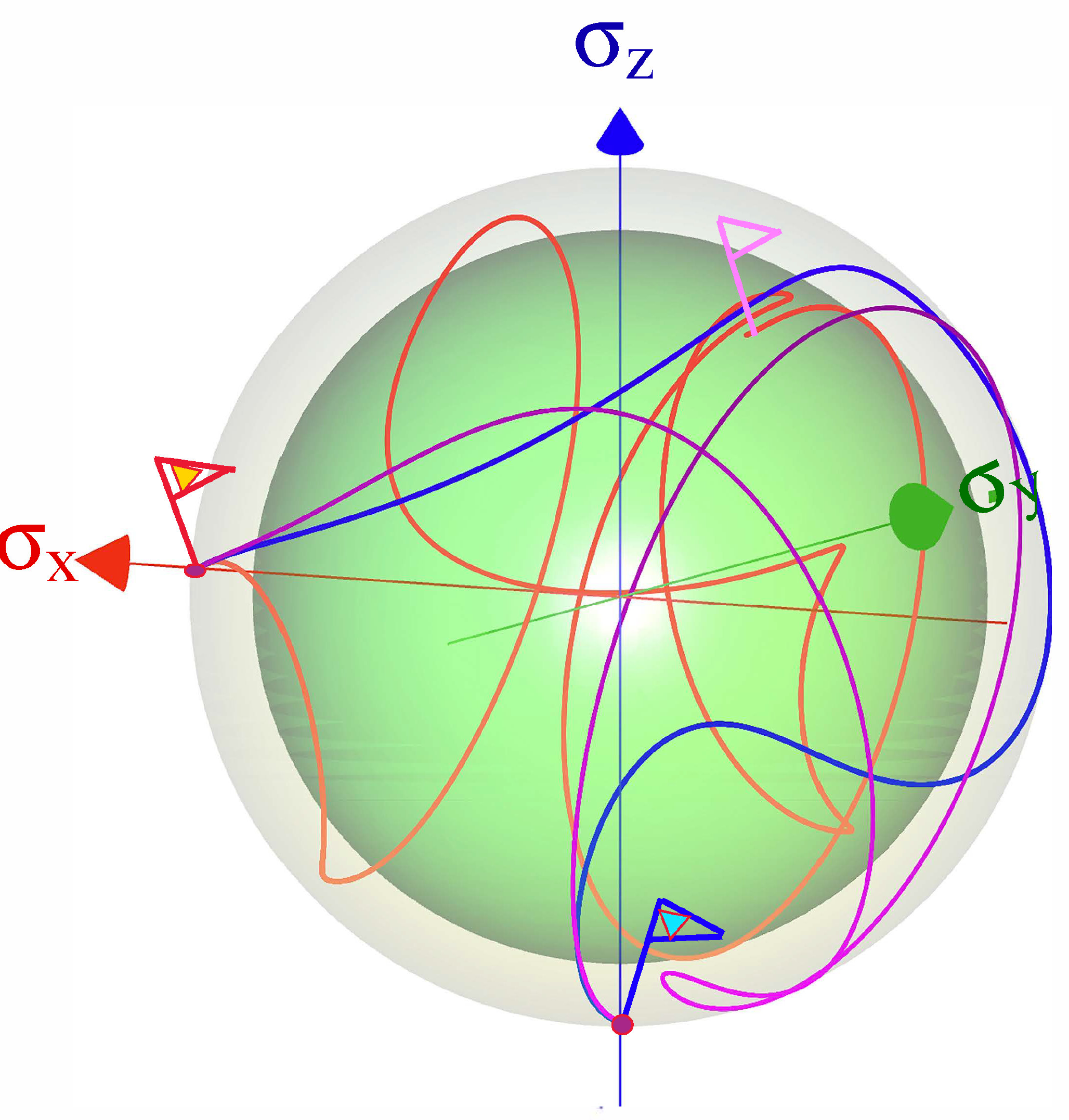}
    \caption{Control trajectory for the Hadamard transformation: Displaying the transition from the $x$ direction to the $-z$ direction
    (indicated by flags). The qubit state under unitary  control with no environmental coupling is represented by the blue curve, while
    the purple line depicts the optimal state trajectory which is obtained from the control protocol that includes the environmental influence (corrected protocol). The red trajectory corresponds to the state dynamics under a unitary field which does not account for the environmental influence on the open system. Overall, we find that the optimal trajectory (purple)
    resides on the surface of the Bloch sphere. Trajectories entering the  inner ball represent loss of purity associated with uncorrected transformation (the final state is indicated by a pink flag).  Other orthogonal directions show a similar pattern.} 
    \label{fig:hadamard-ball-2}
\end{figure}

Figure \ref{fig:hadamard-ball-2} compares the dynamical map  generated trajectories associated with the cases (a),(b) and (c). Clearly, the trajectory of the unitary control protocol under noise (procedure b) misses the target, while the isolated dynamics (a) and the optimized open system protocol (c) leads to the desired final state. The trajectory is close to the surface of the sphere and therefore  is close  to a unitary path. The possible mechanism resemble decoherence control by tracking \cite{katz2007decoherence}.     %a unitary transformation without noise, a transformation with significant dissipation 
    %which is not able to reach the target, and finally a fresh optimization able to correct for the dissipation. 
    %We found corrected trajectories
    %that follow the unitary path.
    %These solutions resemble decoherence control by tracking \cite{katz2007decoherence}.

Overall, we find the expected result,  the control precision degrades when the dissipation increases.  This can be observed in Fig. \ref{fig:unitfid}
where the control objective was studied with increased system bath coupling.
The fidelity is nevertheless
at least an order of magnitude better compared
to the uncorrected control protocol.
The  two-qubit-gate optimizations show
a similar behaviour.

%To obtain a broader perspective regarding the robustness of the model, we present the deviance of the objective functional as a function of natural lifetime of the system in figure \ref{fig:unitfid}. As expected, the increase of the system-environment interaction decreases the ability to execute the transformation to high precision. 
%Nevertheless,  in comparison to the uncontrolled reference, the error retrieval of the control increases with the coupling. Therefore, under relatively strong coupling the method can keep the fidelity within acceptable values for a feasible transformation. 
Note that even for the strongest coupling presented, the weak coupling condition is still maintained and the typical time to achieve a thermal equilibrium is larger by three orders of magnitude relative to the transformation time.

\subsection{Two Qubit Gates}
\label{subsec:two-qubit}

A universal set of quantum gates can be obtained by adding an entangling gate to the single qubit rotation gates. %Eq. (\ref{eq:rotations_single}).
We demonstrate this task by employing the following drift Hamiltonian
\begin{eqnarray}
{{\hat H}}_S^0 = \omega_1 \sigma_1^z + \omega_2 \sigma_2^z ~~=~~
\left(
\begin{array}{cccc}
-\omega_1-\omega_2   & 0 & 0 & 0\\
0 & \omega_1-\omega_2 & 0 & 0 \\
0 & 0 & -\omega_1+\omega_2 & 0 \\
0 & 0 & 0 & \omega_1+\omega_2 
\end{array}
\right)~~,
\label{h02qb}
\end{eqnarray}
and control term
\begin{equation}
{{\hat V}}(t) = \epsilon(t)( {{\hat \sigma}}_1^{+}{{\hat \sigma}}^{-}_2 + {{\hat \sigma}^{+}}_2{{\hat \sigma}}^{-}_1) = \epsilon(t)
\left(
\begin{array}{cccc}
1   & 0 & 0 & 0\\
0 & 0 & 1 & 0 \\
0 & 1 & 0 & 0 \\
0 & 0 & 0 & 1 
\end{array}
\right)~~.
\label{hC2qb}
\end{equation}
The entangling two qubit transformation is taken to be the square root of the swap gate
\begin{equation}
\Lambda_{\sqrt{S}} = 
\left(
\begin{array}{cccc}
1   & 0 & 0 & 0\\
0 & {1+i}\over{2} & {1-i}\over{2} & 0 \\
0 & {1-i}\over{2} & {1+i}\over{2} & 0 \\
0 & 0 & 0 & 1 
\end{array}
\right)~~.
\label{lamhs2qb}
\end{equation}
In isolated conditions this transformation addresses only the two qubit subspace $\left| 01\right\rangle$ and  $\left| 10\right\rangle$. The transformation then 
becomes a rotation in the $SU(2)$ sub-algebra of
the four levels algebra $U(4)$. In the dissipative
case the control has to minimize population leakage
to other states.
%However, additional coupling of the system to the external environment 
%(carried here by the operator $\hat J_y$ as in the previous sections)  
%induces thermalization so that all the four qubit states reach their correct thermal energetic occupation \textbf{???Not sure what you mean here???}. 
The optimization was performed by a similar method as described in subsection \ref{subsec:maps}.

% SK: This seems to be wrong... I am not sure why: Note that for the degenerate case $\omega_2 = \omega_1 $ the subspace of the two entangled states contains only a $\sigma_x$ and is therefore uncontrollable in the field free regime. probably the comutation relations CAN generate the needed operators even with degeneracy.

\begin{figure}
    \centering
    \includegraphics[width=12cm]{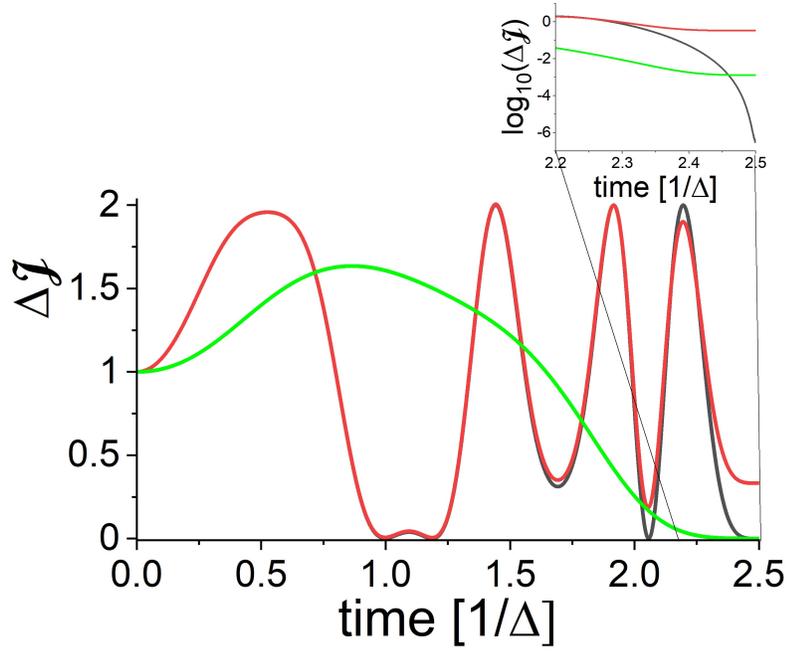}
    \caption{Control to obtain the two-qubit  square root swap  gate,   (\ref{lamhs2qb}): Infidelity as a function of time designation similar to fig. \ref{fig:unit}. %The objective is related to the trace distance $\cal F$ (fidelity) can be related to the objection ${\cal{F}}={\cal J}/4$.
    For the demonstration we chose $\omega_2 = 1.1 \omega_1 $, with $ \omega_1= \Delta = 3\times 10^{-3}$ a.u..}
    \label{fig:2qbunit}
\end{figure}

The control protocol has been studied by employing the same three schemes of Sec. \ref{subsec:i-maps}.
%Namely, the optimization were carried in the isolated regime. Then, to set the reference for the control, the coupling to the environment was turned on, and the degrading of the isolated solution was observed. Finally, a full optimization was performed to examine the capabilities of the quantum control to protect from thermal degrading.
Fig. \ref{fig:2qbunit}, displays the objective functional  ${\cal J}$ as a function of time. 
The figure's inset shows the deviation of the objective from the target state in a logarithmic scale, during the final protocol stage. Uncorrected for the environmental influence within the control, the system deviates considerably from the objective, while a complete optimization, including the environmental, reaches the objective with high fidelity.

%The transformation is characterized by an increase in entropy increase, showing a similar trend as the single qubit gate, see Fig. \ref{fig:2qbunit1}.    
\begin{figure}
    \centering
    \includegraphics[width=9cm]{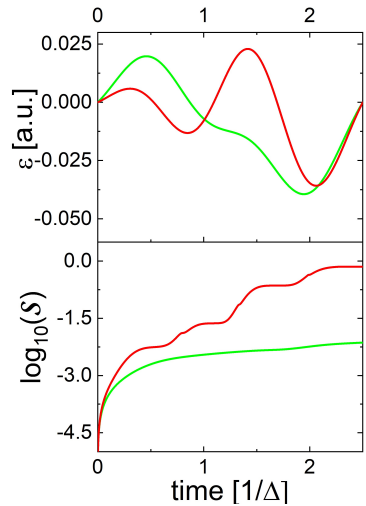}
    \caption{Two qubit gate: (lower) Entropy as a function of time. (upper) The optimal field as a function of time. Designation similar to Fig.
    \ref{fig:unit1}.}
    \label{fig:2qbunit1}
\end{figure}

The control trajectories 
can be graphically depicted by evaluating the operators of the $SU(2)$ algebra, and characterized by the generalized purity. This measure is defined as the purity of the projected state on the $SU(2)$ algebra\cite{barnum2003generalizations}.
Employing such a representation, a similar picture to  Fig. \ref{fig:hadamard-ball-2} emerges. The  successful gates maintain  constant generalised-purity, while the uncontrolled ones degrade the generalized-purity as a result of the coupling to the environment.

\section{Discussion and Summary}
\label{eq:disciss}

The presented analysis constitutes a thermodynamical consistent model that enables quantum control of open systems. The theory was demonstrated by studying 
entropy changing state to state transformations and unitary gates under external influence. 
Previous work, both experimental and theoretical, have addressed  optimal control for cooling transformations, under the condition that the unitary (control) and dissipative parts are independent. \cite{bartana2001laser,aroch2018optimizing,stollenwerk2020cooling,viteau2008optical}.
%Moreover, these studies assumed that the dissipative part was independent of the control protocol.
Such an assumption ignores the dressing of the system by the field and may violate the laws of thermodynamics \cite{levy2014local,hofer2017markovian}.
By building upon a complete description of the total system (including the field), this discrepancy was fixed and employed to achieve control in the present analysis. 
In addition, our examples were carried under 
weak dissipation conditions, achieving control objectives in conditions where unitary dynamics  cannot reach.

From a control theory perspective we can differentiate 
between two types of control objectives, state-to-sate transformations and dynamical maps. For a state-to-state transformation of open systems a controlablility criterion was defined in Ref. \cite{dann2020fast,dirr2019}. Our state to state control  tasks comply with these criteria and are therefore controllable. In accordance with the theory, the CRAB like random optimization achieved the target state with high fidelity. 
Similarly, for isolated systems  a controlability criterion has  been stated for unitary maps
\cite{d2007introduction,huang1983controllability}. However, a controlability theorem for open system maps does not exist. Some progress in this direction has been achieved by a recent study which addressed the adiabatic reset problem \cite{venuti2021optimal} from an optimal control perspective. In addition, 
the ability to perform unitary gates under noise resembles ideas from dynamical decoupling \cite{viola1999dynamical,basilewitsch2021engineering}.
The present results serve as a computational demonstration that practical control of gates under dissipative conditions is possible.

A thermodynamical analysis reveals a new paradigm for the control mechanism. All the control protocols 
are accompanied by significant entropy production.
The intuitive expectation that unitary controls exists in a decoherence free subspace \cite{lidar1998decoherence} is contrary to the observation of very large dissipation. Notably, 
for unitary targets the control trajectory maintains 
high purity along its path, while the state remains far from the instantaneous attractor, implying large entropy production. This is the hallmark of active cooling.

The obtained protocols can be incorporated within a variety of technological procedures. For example,
a standard quantum computation, based on the quantum circuit model, requires an initial pure state and the ability to perform unitary transformation accurately. In practice, there always exists a classical uncertainty in the initial state due to finite temperature of the  environment. In addition, the idealized quantum gates are subject to external noise, inducing an undesired non-unitary evolution on the qubits. The presented control scheme addresses both problems. $\Lambda_R$ incorporates the environmental influence in the resetting process, allowing to accurately prepare the quantum register in the desired initial state. Moreover, the single qubit rotation maps $\Lambda_O$ as well as the two qubit entangling gate $\Lambda_{\sqrt{S}}$ take the dissipation into account. This enables achieving unitary transformations with improved fidelity.
Utilizing such control in noisy quantum information processing can potentially boost their performance.

Alternatively, utilizing the present control scheme, one can also induce controlled non-unitary operations. These can be incorporated in the realization of non-unitary quantum computations \cite{terashima2005nonunitary,mazzola2019nonunitary}.  Finally, the ability to generate a directly controlled entropy change can pave the way to new cooling (and maybe heating) mechanisms, a research field that has been under  extensive attention during the last two decades \cite{goerz2014optimal,koch2016controlling,basilewitsch2019reservoir,dann2019shortcut,dupays2019shortcuts,dann2020fast,alipour2020shortcuts,schirmer2002quantum}.

The capabilities of the model were shown to allow both the generation of non-unitary transformation, as well as an efficient generation of unitary transformation under similar dissipative condition. 
This work paves the way for multiple interesting directions that could be explored in the future. For example, the investigation of open systems quantum speed limits, the application of formal control methods, the inclusion of non-Markovian effects, and the embedding of quantum control in the framework of thermodynamics. These, and others, might leads to new insights that could improve our ability to understand and apply control in the quantum world.

\section*{Acknowledgement}
 We thank Nina Megier for insightful discussions. This study was supported by the Adams Fellowship  Program of the Israel Academy of Sciences and Humanities and by the Israel Science Foundation (Grants No. 510/17 and 526/21).

\appendix
\renewcommand\thefigure{\thesection.\arabic{figure}}
\section{Control Master equation}
\label{appendix-A}
Consider a driven quantum system interacting with an external environment. 
The time dependent Hamiltonian is of the form 
\begin{equation}
\label{eq:ham_tot}
    \hat{H} = \hat{H}_S\b t+\hat{H}_E+\hat{H}_I~~,
\end{equation}
where $\hat{H}_S$ and $\hat{H}_E$ are the bare system and environment Hamiltonian and $\hat{H}_I$ is the system-environment coupling term. 
We consider an interaction term with the following form 
\begin{equation}
    \hat{H}_I = \hat{S}\otimes \hat{B}
\end{equation}
where $\hat{S}$ and $\hat{B}$ are Hermitian operators of the system and environment. Such an interaction is chosen in order to simplify the presentation, the generalization to multiple interaction terms follows the same procedure.

Our present goal is to derive an effective equation of motion for the system, influenced by the external degrees of freedom. 
Applying the Born-Markov approximation the reduced dynamics in the interaction picture relative to the free dynamics are governed by the Quantum Markovian Master Equation \cite{breuer2002theory}
\begin{equation}
    \f d{dt}\tilde{\rho}_{S}\b t=-\f 1{\hbar^{2}}\int_ 0^{\infty}{ds\,{\rm{tr}}_{E}\b{\sb{\tilde{H}_{I}\b t,\sb{\tilde{H}_{I}\b{t-s},\tilde{\rho}_{S}\b t\otimes\hat{\rho}_{E}}}}}~~.
    \label{eq:25n}
\end{equation}
These approximations assume weak system-environment coupling and a rapid decay of the environmental correlations. In addition,  the environment is negligibly affected by the interaction with the system, meaning it remains stationary throughout the dynamics $\hat{\rho}_E \equiv\hat{\rho}_E\b 0$. 
We expand the interaction Hamiltonian in terms of system's eigenoperators 
\begin{eqnarray}
    \tilde{H}_I =&\hat{U}^\dagger\b t\hat{H}_I \hat{U}\b t\nonumber\\
    =&\hat{U}_S^\dagger\b t\hat{S}\hat{U}_S\b t\otimes\hat{U}_E^\dagger \b t \hat{B}\hat{U}\b t \nonumber\\=&
    \sum_k c_k\b t \hat{U}_S^\dagger\b t\hat{F}_k\b t\hat{U}_S\b t\otimes\tilde{B}\b t \label{eq:26n}\\=& 
    \sum_k c_k\b t \hat{F}_k\b t e^{-i\theta_k\b t}\otimes\tilde{B}\b t\nonumber
    \\=& 
    \sum_k \eta_k\b t \hat{F}_k\b t e^{-i\Lambda_k\b t}\otimes\tilde{B}\b t\nonumber~~,
    \\=&
    \sum_k \eta_k\b t \hat{F}_k^\dagger\b t e^{i\Lambda_k\b t}\otimes\tilde{B}^\dagger\b t\nonumber~~
\end{eqnarray}
where $c_k=\eta_k\b t e^{-i\lam_k\b t}$ are expansion coefficients of $\hat S$ in terms of the eigenoperators $\{\hat{F}_k\b t\}$, $\eta_k,\lam_k\in \mathrm{R}$ and 
$\Lambda_k\b t=\theta_k\b t +\lam_k\b t$. In the last equality we utilized the Hermitiacy of $\hat{S}$ and $\hat{B}$.
Next, we substitute Eq. (\ref{eq:26n}) into (\ref{eq:25n}) to obtain 
\begin{eqnarray}
    \f d{dt}\tilde{\rho}_{S}\b t=\f 1{\hbar^{2}}\int_ 0^{\infty}{ds\,{\rm{tr}}_{E}\b{\sb{\tilde{H}_{I}\b{t-s}\tilde{\rho}_{S}\b t\hat{\rho}_{E}\tilde{H}_{I}\b t-\tilde{H}_{I}\b t\tilde{H}_{I}\b{t-s}\tilde{\rho}_{S}\b t\hat{\rho}_{E}}}}+{\rm{h.c}}
       \nonumber\\
       =\f 1{\hbar^{2}}\sum_{ kk'}\int_0^{\infty}ds\,e^{-i\b{\Lambda_{ k}\b{t-s}-\Lambda_{ k'}\b t}}\mean{\tilde{B}\b t\tilde{B}\b{t-s}}_{E}\eta_{k'}\b t\eta_{ k}\b{t-s}  \\\times\sb{
    \hat F_{k}\b{t-s}\tilde{\rho}_{S}\b t\hat F_{k'}^{\dagger}\b t-\hat F_{k'}^{\dagger}\b t
    \hat F_{k}\b{t-s}\tilde{\rho}_{S}\b t +{\rm{h.c}}}\nonumber
\end{eqnarray}
Under Markovian dynamics the environment correlation decay rapidly relative to the intrinsic timescale of the system. Here we also assume that the environment dynamics is much faster then the typical timescale of the drive. Under this condition, the integral is dominated by the value of the integrand in the range $s\in{\sb{0,\tau_E}}$, where $\tau_E$ is typical timescale associated with the decay of  environmental correlations. In this physical regime, the eigenoparators and coefficients do not change much and we can approximate $\hat{F}_k\b{t-s}\approx \hat{F}_k\b t$ and $\eta_k\b{t-s} \approx \eta_k\b t$, leading to
\begin{eqnarray}
        \f d{dt}\tilde{\rho}_{S}\b t =\Xi_{ kk'}\b t\sb{\hat{F}_{k}\b t\tilde{\rho}_{S}\b t\hat F_{k'}^{\dagger}\b t-\hat F_{k'}^{\dagger}\b t\hat F_{k}\b t\tilde{\rho}_{S}\b t}+{\rm{h.c}}~~,
\end{eqnarray}
with
\begin{equation}
    \Xi_{ kk'}\b t=\f 1{\hbar^{2}}\int _0^{\infty}{ds\,e^{-i\b{\Lambda_{ k}\b{t-s}-\Lambda_{ k'}\b t}}\mean{\tilde B\b{s}\tilde B\b{0}}_{E}\eta_{ k'}\b t\eta_{ k}\b t}~~,
    \label{apeq:29}
\end{equation}
where we utilized the invariance of correlation functions under time-translation for a stationary environment.
The rapid decay of correlation also allows expanding the phases near $t$ as $s \simeq \tau_E$: 
\begin{equation}
    \Lambda_k\b{t-s}=\Lambda_k\b t-\Lambda_k\b{t}s\equiv \Lambda_k\b{t}-\omega_k\b t s~~.
    \label{apeq:30}
\end{equation}
Substituting Eq. (\ref{apeq:29}) into (\ref{apeq:30}) we get terms proportional to $e^{-i\b{\Lambda_k\b{t}-\Lambda_k'\b{t}}}$. For $k\neq k'$ these typically rotate rapidly and average out to zero. As a result,  mixed terms the master equation vanish, leading to a Gorini Kossakowski Sudarshan Lindblad (GKLS) form 
\begin{eqnarray}
        \f d{dt}\tilde{\rho}_{S}\b t=-\f i{\hbar}\sb{\hat H_{LS}\b{\omega_k\b t,t},\tilde{\rho}_{S}\b t}\nonumber\\+\sum_{k}\gamma_{k}\b{\omega_k\b t,t}\b{\hat F_{k}\b t\tilde{\rho}_{S}\b t\hat F_{k}^{\dagger}\b t-\f{1}{2}\{\hat F_{k}^{\dagger}\b t\hat F_{k}\b t\tilde{\rho}_{S}\b t\}}~~.
\end{eqnarray}
where
\begin{equation}
    \gamma\b{\omega\b t,t}=\Gamma\b{\omega,t}+\Gamma^{*}\b{\omega,t}=\int_{-\infty}^{\infty}{ds\,e^{i\omega s}\mean{\tilde{B}\b s\tilde{B}_{\beta}\b 0}_{E}}
\end{equation}
and
\begin{equation}
    \hat H_{LS}\b t=\sum_{k}R\b{\omega_{k}\b t}F_{k}^{\dagger}\b tF_{k}\b t
\end{equation}
with
\begin{equation}
    \Gamma\b{\omega_k\b t,t}=\f 1{\hbar^{2}}\eta_{ k}^2\b t\int_0^{\infty}{ds e^{i\omega_k\b t s}\mean{\hat{B}\b s \hat{B}\b 0}_{E}}
\end{equation}
and
\begin{equation}
R\b{\omega,t}=\f 1{2i}\b{\Gamma\b{\omega,t}-\Gamma^{*}\b{\omega,t}}~~.
\end{equation}
Overall, the obtained master equation is valid in the weak coupling limit, assuming a Markovian environment. In addition, the typical timescale characterizing the drive may be fast relative to the system, but should be slow relative to the decay of environmental correlations.

\section{Numerical comments}
\label{appendix-B}
When employing Eq. (\ref{eq:eigenf}) one needs to retrieve the time-dependent phase of $\hat F_j$, the eigenoperators of the propgator ${\cal U}_S(t,0) $.  We employ a numerical diagonalization of ${\cal U}_S(t,0) $  at each time step. We then ensure a continuous labeling of the eigenoperators. This task is achieved by calculating the overlap between the current and updated set of operators. 
A conventional retrieval of the phase with inverse trigonometric functions was found to suffer from erratic behaviour. In order overcome this issue we employed the following procedure: 
For an unknown phase function 
$f(t) = e^{i\theta(t)}$, the formal derivative of the function is given by:
\begin{equation}
{{d} \over{dt}} f(t) = i \dot{\theta} f    \to       \theta =  i\int_0^t{ {\dot{f} dt}\over{f}} 
\end{equation}
so that the phase is retrieved stably by integration.

Another important numerical comment concerns the number of frequency components $M$, in Eq. (\ref{eq:cont_exp}). The number of frequencies required to modify the entropy in a cooling and heating protocol are displayed in Fig. \ref{fig:vs_freq}. Typically, the objective improves with the number of frequencies in the control. However, in the cooling fields, we witness a saturation with the increase of $M$. In addition, due to the increase in complexity achieving the objective becomes increasingly harder when increasing the number of energy levels.
\setcounter{figure}{0}
\begin{figure}
    \centering
    \includegraphics[width=8cm]{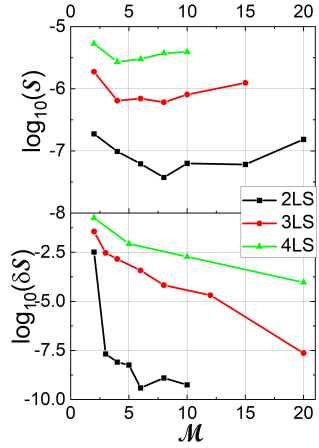}
    \caption{The entropy difference as the objective functional as a function of the number of field frequency components $M$. The control task is compared for two (black), three (red) and four (green) level systems, for cooling (upper panel) and heating (lower panel) processes.} 
    \label{fig:vs_freq}
\end{figure}

\section{Scaling of the objective with the system environment coupling.}

\begin{figure}
    \centering
    \includegraphics[width=11cm]{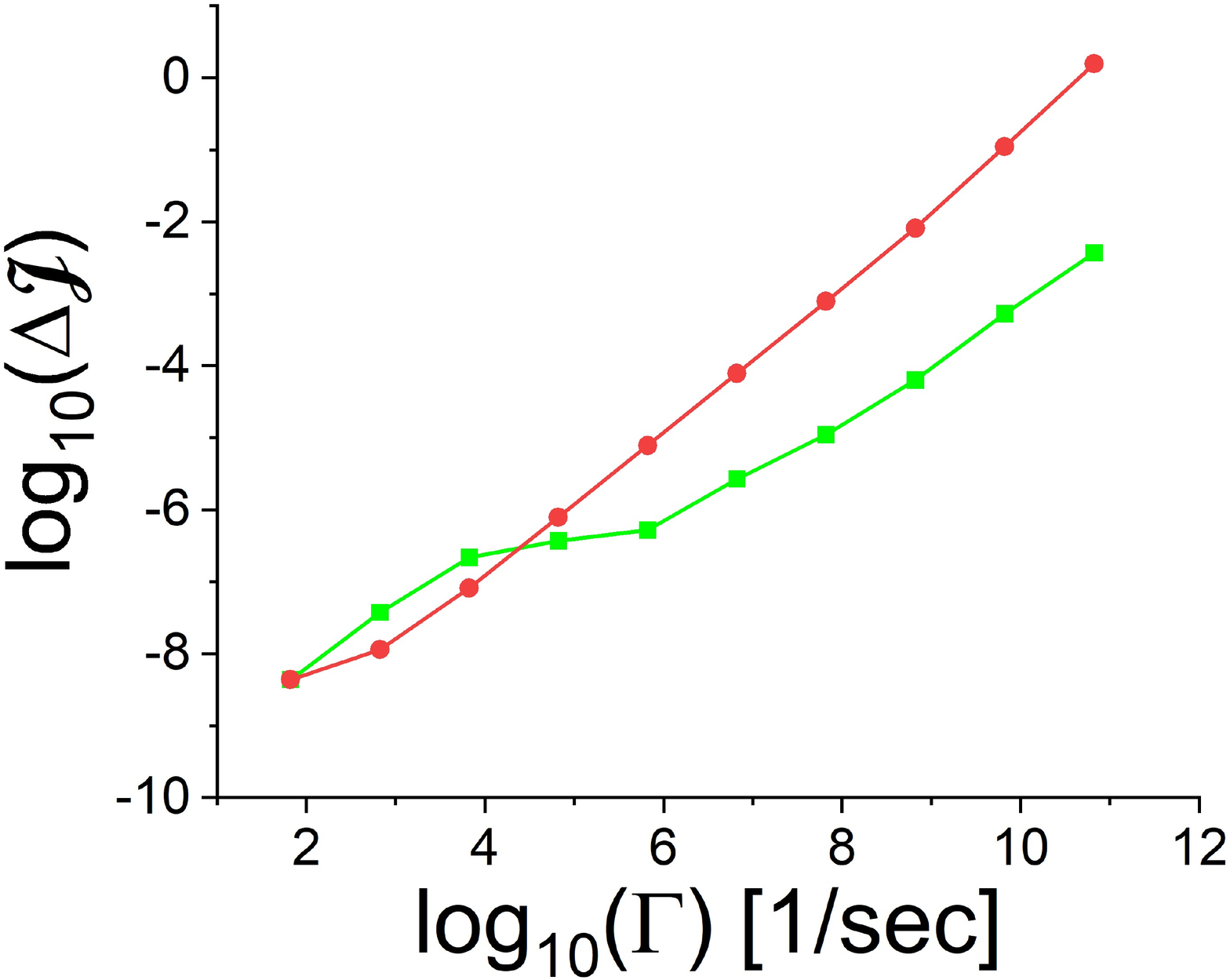}
    \includegraphics[width=11cm]{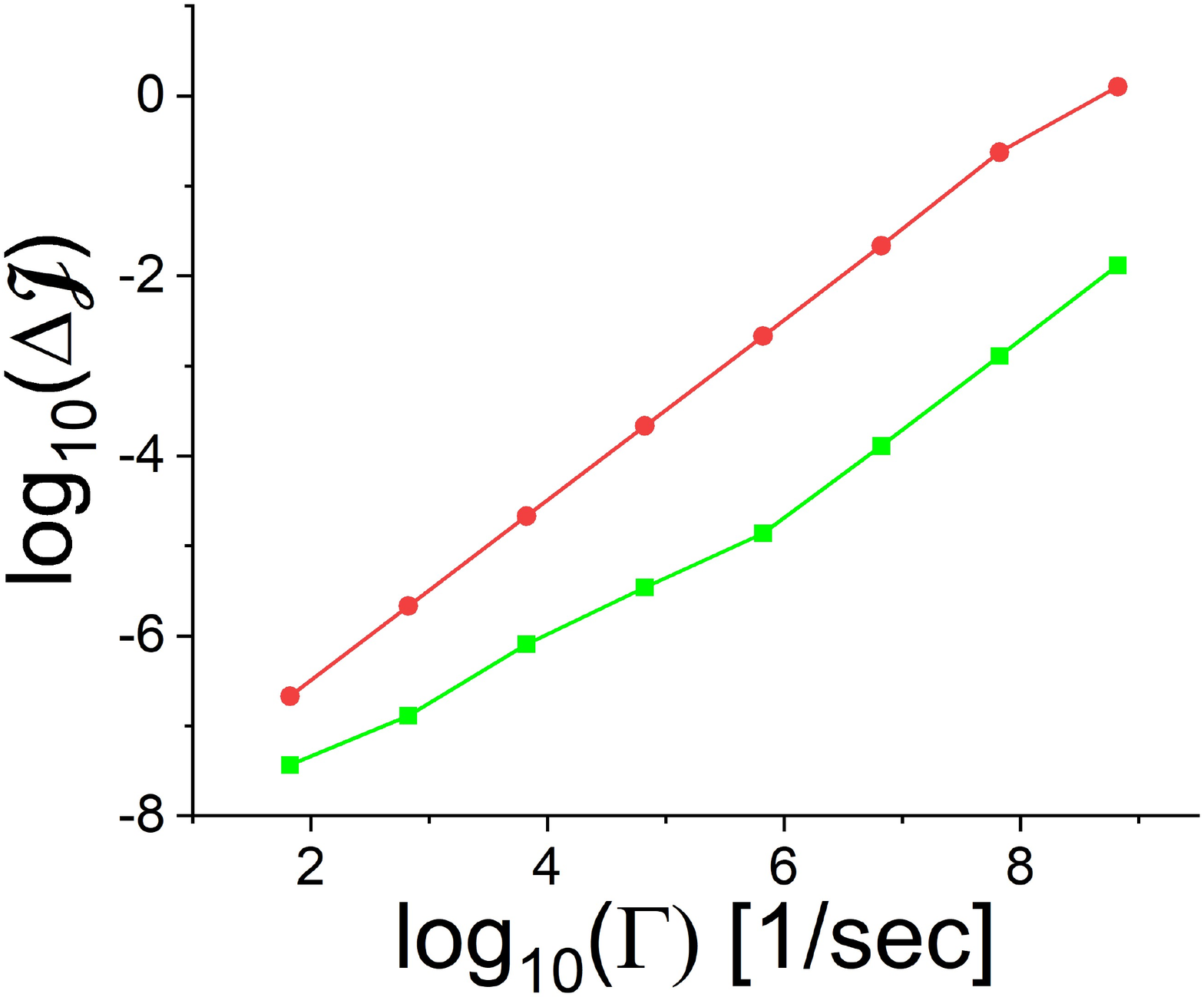}
    \caption{Degradation of a Unitary transformation:
    The precision $\Delta {\cal J}$    in log scale as a function of decay parameter $\Gamma = k_{\downarrow}+k_{\uparrow}$
    for the drift Hamiltonian ($\hat V(t)=0$).
    To Hadamar transformation, bottom two qubit gate.
    (green) The infidelity for the optimal transformation as a function of system-environment coupling. Each point represents the best case optimization computed from scratch for a given coupling strength. (red) The infidelity obtain form the bath free optimization
    when the bath is included. Cf.  Fig. \ref{fig:unit}.}
    \label{fig:unitfid}
\end{figure}

As the system-environment coupling $g$ Eq. (\ref{eq:kinetic_coeff})
increases the control objective decreases.
Figure \ref{fig:unitfid} shows in a log-log plot the degradation of the transformation as a function of the effective decay rate 
$\Gamma=k_{\downarrow}+k_{\uparrow}$, calculated in the absence of driving.   

%\bibliography{scibib}

\bibliographystyle{Science}

\end{document}